\documentclass[12pt]{article}
\usepackage{amsmath,amssymb,bm,epsfig}
\usepackage[subrefformat=parens]{subcaption}
\usepackage{cite}
\usepackage{color}
\renewcommand{\thefootnote}{\fnsymbol{footnote}}

\begin{document}

\title{
\begin{flushright}
\begin{minipage}{0.2\linewidth}
\normalsize
EPHOU-17-005 \\*[50pt]
\end{minipage}
\end{flushright}
{\Large \bf {
Toward pole inflation and attractors in supergravity : Chiral matter field inflation
\\*[20pt]}}}

\author{
Tatsuo~Kobayashi${}^1$ \footnote{
E-mail address:  kobayashi@particle.sci.hokudai.ac.jp}, \ \ 
Osamu~Seto${}^{2,1}$ \footnote{
E-mail address: seto@particle.sci.hokudai.ac.jp}, \ and \
Takuya~H.~Tatsuishi${}^1$ \footnote{
E-mail address: t-h-tatsuishi@particle.sci.hokudai.ac.jp} \\
\\*[20pt]
{\it \normalsize
${}^1$ Department of Physics, Hokkaido University,  Sapporo 060-0810, Japan }\\  
{\it \normalsize
${}^2$ Institute for International Collaboration, Hokkaido University,}\\{\it \normalsize Sapporo 060-0815, Japan }
}

\date{
\centerline{\small \bf Abstract}
\begin{minipage}{0.9\linewidth}
\medskip 
\medskip 
\small
In string-inspired supergravity theory, K\"{a}hler metric of chiral matter fields often has a pole.
Such K\"{a}hler metric is interesting from the viewpoint of the framework of the pole inflation, where the scalar potential can be stretched out to be flat around the pole for a canonically normalized field and inflation can be realized.
However, when K\"{a}hler metric has a pole, the scalar potential can also have a pole at the same point in 
supergravity theory.
We study such supergravity models with the pole, and provide numerical analysis of inflationary dynamics and resultant density perturbation.
In contrast with usual pole inflation models, inflation in this supergravity based model occurs
 not on the pole but region apart from the pole.
We show that the existence of the pole in the scalar potential is crucial nevertheless.
We also examine attractor behavior of our model.
\end{minipage}
}

\begin{titlepage}
\maketitle
\thispagestyle{empty}
\clearpage
\thispagestyle{empty}
\end{titlepage}

\renewcommand{\thefootnote}{\arabic{footnote}}
\setcounter{footnote}{0}
\vspace{35pt}
\section{Introduction}

Exponentially accelerated expansion of spacetime at the very early Universe, so called inflation,
 solves various problems in the standard Big Bang cosmology~\cite{Guth:1980zm}.
Nowadays the inflationary cosmological model is the standard paradigm in cosmology.
In most of early studies~\cite{Guth:1980zm,Sato:1980yn,Linde:1981mu,Albrecht:1982wi}, 
 it had been thought that the Higgs field in a Grand Unified Theory
 provides a false vacuum energy for inflationary expansion.
Since it had been pointed out that
 any gauge singlet scalar field with a chaotic initial condition can derive inflationary cosmic expansion
 if its scalar potential is flat enough~\cite{Linde:1983gd},
 a scalar field to drive inflation, called inflaton $\phi$, has been thought
 to be not necessarily a Higgs field related with any gauge theory.

When we refer recent observational results of temperature fluctuation
 in the Cosmic Microwave Background (CMB)~\cite{Ade:2015xua,Ade:2015tva},
 so-called Higgs inflation where the Higgs field in the standard model of particle physics
 with nonminimal coupling to gravity plays a role of inflaton fits the observational data well~\cite{Bezrukov:2007ep}.
The secret of the success of Higgs inflation is the Weyl rescaling from the Jordan frame to the Einstein frame,
 the exponential function factor for this frame transformation
 makes the scalar potential in the Einstein frame $V_E$ for the canonical normalized inflaton $\chi$
 being exponentially flatter as
\begin{equation}
 V_E(\chi) \sim \left( 1 - e^{-\frac{2\chi}{\sqrt{6}M_P}} \right)^{2} , 
\end{equation}
 with $M_P$ being the reduced Planck mass.
An exponential-like potential can particularly well fit the observational data,
 because for such a model there is an interesting relation
\begin{equation}
 n_s \simeq 1 - \frac{2}{N_e}, 
\end{equation}
 where $n_s$ and $N_e$ denote the spectral index of its density perturbation and the number of e-folds, respectively.
Then, for a sensible number of e-folds $N_e \simeq 60$,
 the best fit value of $n_s \simeq 0.97$ can be automatically reproduced.
This property can been also seen in the so-called Starobinsky inflation model proposed back in 1980,
 where inflaton is a scalar degrees of freedom appears through the Weyl rescaling and, as the common result,
 has an exponential dependent scalar potential~\cite{Starobinsky:1980te}.
Hence, as is well known, 
 both Higgs inflation and Starobinsky inflation well explain observational results. 

The generality of this feature with an exponential function in the scalar potential
 can be classified and be well understood in the framework of $\alpha$-attractor model,
 where the scalar potential is expressed as
\begin{equation}
 V_E(\chi) \sim \left( 1 - e^{-\sqrt{\frac{2}{3\alpha}}\frac{\chi}{M_P}} \right)^2 , 
\end{equation}
 with new model parameter $\alpha$~\cite{Kallosh:2013hoa,Ferrara:2013rsa,Kallosh:2013tua,Kallosh:2013yoa}. 
The original Starobinsky inflation model corresponds to $\alpha=1$. 
In Ref.~\cite{Kallosh:2013yoa}, the attractor behavior of the parameter $\alpha$ has been shown.
Since it was recognized that the essence of above key feature comes from the fact that
 the kinetic term in the original field or the original frame has a singular pole, 
 this classification of models was extended and
 those models have been regarded as pole inflation model~\cite{Galante:2014ifa,Broy:2015qna}.
If a field has a singularity in the kinetic term,
 the field is infinitely stretched around the singular point of the kinetic term by canonical normalization.
In this case that the scalar potential has no singularity at the kinetic singular point,
 the scalar potential is also stretched infinitely.
Then, we can realize sufficiently flat potential, where the inflation by the canonical field takes place. 
Thus, the pole inflation is interesting,
 and some extensions also have been studied (see e.g. Ref.~\cite{Terada:2016nqg,Nakayama:2016eqv}).

Indeed, the K\"{a}hler metric of chiral matter fields can often have a pole within the framework
 of string-inspired supergravity theory, which typically leads to the following form of kinetic term,
\begin{equation}
  \mathcal{L}_{KE}=K_{\phi\bar{\phi}}|\partial_\mu\phi|^2=\frac{p}{(1-|\phi|^2)^2}|\partial_\mu\phi|^2 ,
\label{eq:kinetic term_phi}
\end{equation}
 where $K_{\phi\bar{\phi}}$ is the K\"{a}hler metric and $p$ is a real number.
This kinetic term has a singular point at $\phi=1$.
Here we use the unit with the Planck scale $M_{P}=1$.
The above K\"{a}hler metric can be obtained from the K\"{a}hler potential $K=-p\ln(1-|\phi|^2)$.
Thus, the supergravity model with this K\"{a}hler potential would be a good candidate for realizing the pole inflation.
Starobinsky inflation models in no-scale supergravity have been studied, for example, in
 Refs.~\cite{Ketov:2010qz,Ketov:2012se,Ellis:2013xoa,Ellis:2014rxa,Kamada:2014gma,Ellis:2014opa,Lahanas:2015jwa,Garg:2015mra,Ellis:2015xna,Romao:2017uwa}. 
However, the corresponding scalar potential can have a similar pole at the same point,
 because the F-term scalar potential is given by
\begin{equation}
 V=e^K\left[K^{I\bar{J}}(D_I W)(D_{\bar{J}}\overline{W})-3\left|W\right|^2\right], 
\label{eq:scalar potential in SUGRA}
\end{equation}
 by K\"{a}hler potential $K$ and superpotential $W$, with $D_I W \equiv K_IW + W_I$,
 and the overall factor $e^K$ is singular at the same point as the kinetic term.
In fact, inflation can not take place on the pole of kinetic term.
That is, the realization of the inflation potential is not trivial within
 the framework of supergravity models with the singular kinetic term.

In this paper, motivated by the fact that the kinetic term of Eq.~(\ref{eq:kinetic term_phi}) often appear
 through a compactification of higher dimensional stringy supergravity models,
 we consider inflationary potentials consist of
 F-term scalar potential (\ref{eq:scalar potential in SUGRA}) with a logarithmic K\"{a}hler potential, 
 constant superpotential and additional non-vanishing F-term component by another field,
 and a constant (or sufficiently flat) term in the scalar potential.
We find that the potential has a flater region apart from the pole
 and, by use of it, the inflarionary expansion can be realized.
Unlike the original pole inflation model~\cite{Galante:2014ifa,Broy:2015qna},
 our inflation occurs at regions apart from the pole of the kinetic term 
 because the scalar potemtial also has the pole at the same point.
Nevertheless the existence of the pole is important to realize flat region in the potential in our model.
We show examples of an inflationary potential construction
 under the situation that both the kinetic term and the scalar potential have the pole at the same place,
 which is a case in a class of supergravity.

This paper is organized as follows.
In section 2, we explain the structure of scalar potential in our model. 
In section~\ref{sec:constant potential}, at first, 
 we show numerical analysis of inflationary observables and 
 attractor behavior of parameter dependence
 in a model with $\phi$-independent superpotential and uplifting term.
In section~\ref{sec:superpotential}, we include inflaton-dependence in our superpotential
 and show that it does not spoil succesful inflationary expansion.
In section~\ref{sec:D-term potential},
 we consider a case of $\phi$ dependent uplifing term.
Section~\ref{sec:Conclusion} is devoted to conclusions and discussions.

\section{Inflation potential}
\label{sec:}

In superstring theory, the following type of K\"{a}hler potential
\begin{equation}
 K=-p\ln(T+\bar{T}-A|\phi|^2) ,
\end{equation}
appears in the consequence of compactification of extra dimensions,
where $T$ is a K\"{a}hler modulus, $\phi$ is a chiral superfield, and $A$ is a constant coefficient.
Here we use the notation that the superfield and its lowest component are denoted by the same letter.
We assume the K\"{a}hler modulus $T$ to be stabilized with a sufficiently heavy mass by a certain mechanism, 
and replace $T$ with its vacuum expectation value $\langle T \rangle$ below such a mass scale.
Then, by rescaling  $\phi$, we obtain the following K\"{a}hler potential
\begin{equation}
 K=-p\ln(1-|\phi|^2). \label{eq:Kahler potential}
\end{equation}
The kinetic term of the $\phi$ is given by Eq.~(\ref{eq:kinetic term_phi}), and which has a pole at $|\phi|=1$.
The F-term scalar potential in the framework of supergravity is given by Eq.~(\ref{eq:scalar potential in SUGRA}).

\begin{figure}[htbp]
 \centering
 \begin{tabular}{c}
  \begin{minipage}{0.5\hsize}
   \centering
   \includegraphics[width=70mm]{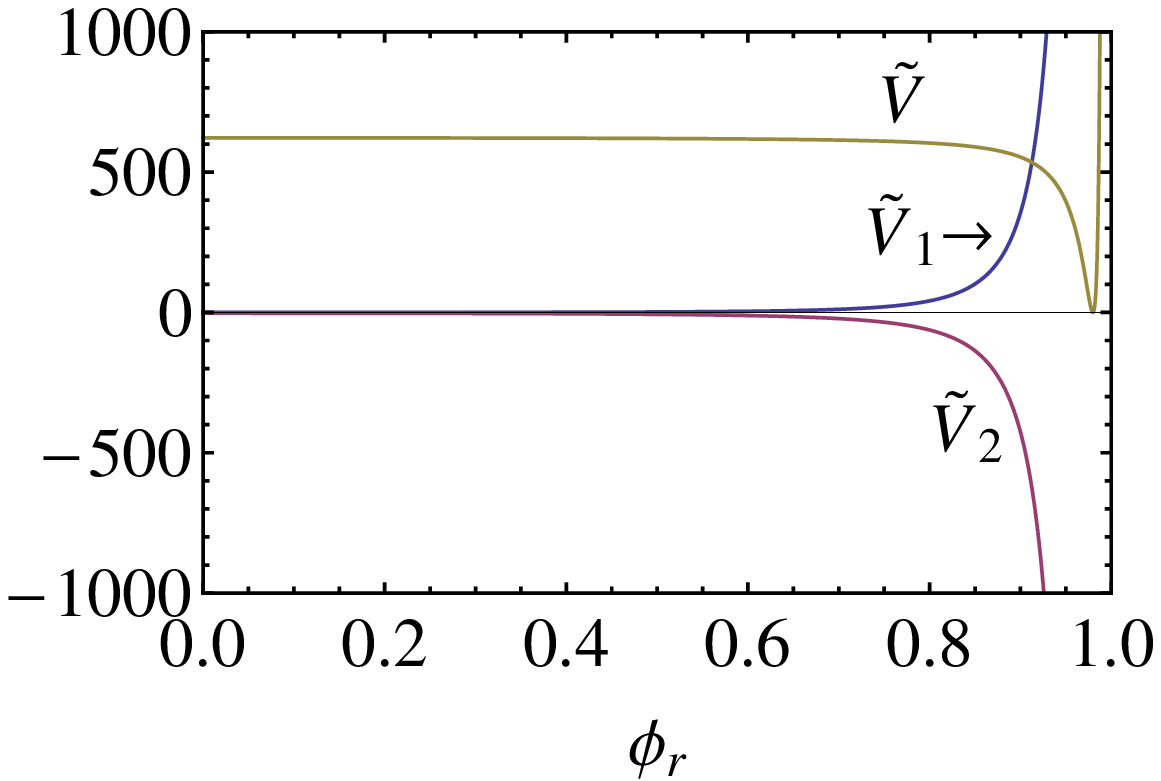}
   \subcaption{} 
   \label{fig:potential_sample_phi}
  \end{minipage}
  \begin{minipage}{0.5\hsize}
   \centering
   \includegraphics[width=70mm]{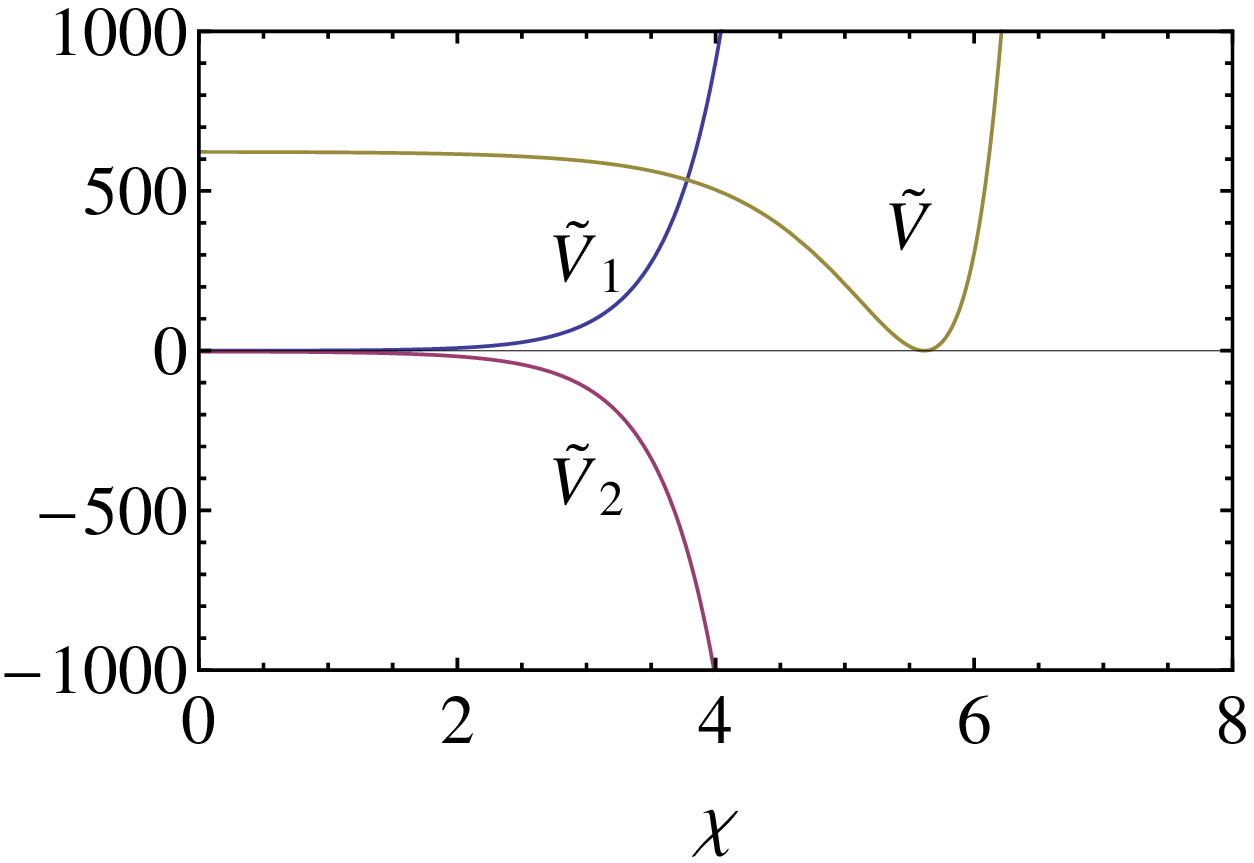}
   \subcaption{} 
   \label{fig:potential_sample_chi}
  \end{minipage}
 \end{tabular}
 \caption{
The shape of potential (\ref{eq:scalar potential_2}) and (\ref{eq:scalar potential_3}) with $p=3,\,\tilde{F}^2=0.08$, and $\tilde{V}_C=625$.
The pole at $\phi_r=1$ in Fig.~\ref{fig:potential_sample_phi} corresponds to the pole at $\chi=\infty$ in Fig.~\ref{fig:potential_sample_chi}.
}
 \label{fig:potential_sample}
\end{figure}

In this section, we assume non-vanishing F-term $|F|^2\neq0$ of other fields and
 a constant potential $V_C$ due to a D-term scalar potential and/or explicit supersymmetry breaking effects,
 and hence we obtain the following scalar potential
\begin{equation}
 V=e^K\left[K^{\phi\bar{\phi}}\left|D_\phi W\right|^2-3\left|W\right|^2+|F|^2\right]+V_C.
\label{eq:scalar potential_1}
\end{equation}
At first, for simplicity, we consider the $\phi$-independent superpotential $W=W_0$.
In the following analysis, we often use the dimensionless scalar potential
\begin{equation}
 \tilde{V}\equiv \frac{V}{|W_0|^2}=\frac{1}{(1-|\phi|^2)^p}\left(p|\phi|^2-3+\tilde{F}^2\right)+\tilde{V}_C,
\label{eq:scalar potential_2}
\end{equation}
 with $\tilde{F}^2=|F|^2/|W_0|^2$ and $\tilde{V}_C=V_C/|W_0|^2$, 
 normalizing the scalar potential (\ref{eq:scalar potential_1}) by $|W_0|^2$.
We also define 
\begin{equation}
\tilde{V}_1=e^K\left(K^{\phi\bar{\phi}}|D_\phi W|^2\right)/|W_0|^2, \qquad \tilde{V}_2=e^K\left(-3|W|^2+|F|^2\right)/|W_0|^2.
\end{equation}
That is, $\tilde{V}=\tilde{V}_1+\tilde{V}_2+\tilde{V}_C$.
Since we have assumed a $\phi$-independent superpotential,
 the scalar potential depends only on the radial component\footnote{
When $\phi_r$-direction is stabilized and $\theta$-direction is slightly sloped, helical phase inflation due to $\theta$ can be occur  \cite{Li:2014vpa,Li:2015taa,Ketov:2015tpa}.}
 of $\phi$, $\phi_r$, hence, without loss of generality, we can identify the inflaton direction with $\phi_r$ which is sometime simply denoted as $\phi$.
The form of potential is shown
 in Fig.~\ref{fig:potential_sample_phi} with special values of $\tilde{F}^2$ and $\tilde{V}_C$. 
The values of $\tilde{F}^2$ and $\tilde{V}_C$ are chosen in such a way that
 the minimum lies between $0<\phi<1$ and its vacuum energy there vanishes. 
Since $\tilde{V}_1$ decreases and $\tilde{V}_2$ increases divergently toward the pole at $\phi_r=1$,
 the potential around the origin tends to be flat compared with the depth of the potential at the stationary point.
(Thus, there is an approximate shift symmetry for small $\phi_r$ region.)

A canonically normalized field $\chi$ which has canonical kinetic energy $\mathcal{L}_{KE}=\frac{1}{2}(\partial_\mu\chi)^2$ can be defined by
\begin{align}
 \phi=\tanh\left(\frac{\chi}{\sqrt{2p}}\right).
\end{align}
The pole $\phi_r =1$ corresponds to $\chi = \infty$.
In terms of the $\chi$, the potential (\ref{eq:scalar potential_2}) is expressed by
\begin{equation}
 \tilde{V}=\cosh^{2p}\left(\frac{\chi}{\sqrt{2p}}\right)\left[p\tanh^2\left(\frac{\chi}{\sqrt{2p}}\right)-3+\tilde{F}^2\right]+\tilde{V}_C,
\label{eq:scalar potential_3}
\end{equation}
and is shown in Fig.~\ref{fig:potential_sample_chi}.
The potential of $\chi$ is not flat around the pole, but singular.
This is because of the factor $e^K$ in the supergravity scalar potential. 
This is different from the simplest pole inflation, where 
$V(\phi)$ is not singular at the pole of the kinetic term~\cite{Galante:2014ifa}.
However, different singular behaviors between $\tilde V_1$ and $\tilde V_2$ create
 a minimum and make its depth deeper.
Then, the potential for smaller $\chi$ region is still flat compared with the depth of the minimum.
Around such a flat region, we can realize the inflation.

\begin{figure}[htbp]
 \centering
 \includegraphics[width=100mm]{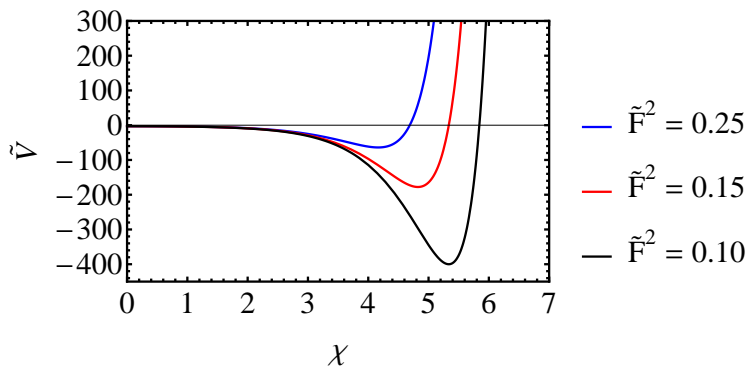}
 \caption{$\tilde{F}$-dependence of the potential $\tilde{V}$ with $p=3$ and $\tilde{V}_C=0$ with respect to the canonical field $\chi$.
}
 \label{fig:delta_dependence}
\end{figure}

The stationary condition of the potential $V=|W_0|^2\tilde{V}$,
\begin{equation}
 V_\chi\equiv\frac{dV}{d\chi}=\frac{d\phi}{d\chi}V_\phi=0 ,
\end{equation}
can be expressed by
\begin{equation}
 \tilde{V}_\phi=\frac{2p\phi[(p-1)\phi^2+\tilde{F}^2-2]}{(1-\phi^2)^{p+1}}=0, \label{eq:Vc_stationary condition} 
\end{equation}
by using the potential (\ref{eq:scalar potential_2}).
A solution of Eq.~(\ref{eq:Vc_stationary condition}) is
\begin{equation}
 \phi^2=\frac{2-\tilde{F}^2}{p-1} \label{eq:Vc_stationary point_chi} ,
\end{equation}
 for $p\neq1$, and the condition for the stationary point to be present at $0<\phi<1$ is
\begin{equation}
 3-p<\tilde{F}^2<2. \label{eq:constraint_F}
\end{equation}
Figure~\ref{fig:delta_dependence} shows the role of $\tilde{F}^2$ in the potential (\ref{eq:scalar potential_3}).
Smaller $\tilde{F}^2$ corresponds to deeper vacuum, and the potential becomes sharp around the minimum.
Hence, as $\tilde{F}^2$ becomes smaller, the potential becomes flatter except around the minimum.

\section{Numerical analysis}
\label{sec:constant potential}

In this section, we study inflation behavior of our potential numerically.

\subsection{Models with $p=2,3$}
\label{sec:models}

The shape of potential of inflaton is restricted by observations of the fluctuation of
 temperature anisotropy in the CMB. 
Slow-roll parameters are defined by
\begin{equation}
 \epsilon=\frac{1}{2}\left(\frac{V_\chi}{V}\right)^2,\quad \eta=\frac{V_{\chi\chi}}{V},
\label{eq:def_slow-roll parameters}
\end{equation}
 and the power spectrum, the spectral index of the power spectrum of density perturbation,
 and the tensor-to-scalar ratio are expressed by
\begin{equation}
 \mathcal{P}_\zeta=\frac{V}{24\pi^2\epsilon},\quad n_s=1+2\eta-6\epsilon,\quad r=16\epsilon,
\label{eq:def_observable quantities}
\end{equation}
respectively.
According to the BICEP2/Keck Array and Planck data~\cite{Ade:2015xua,Ade:2015tva},
 these quantities are measured or bounded as
\begin{equation}
 \mathcal{P}_\zeta= (2.20\pm1.10)\times10^{-9},\quad n_s=0.9655\pm0.0062,\quad r<0.11.
\label{eq:Planck}
\end{equation}
In addition, the number of e-folds
\begin{equation}
 N_e\equiv -\ln\left(\frac{a_\textrm{end}}{a}\right) \simeq \int_{\chi_f}^{\chi_{{}_{N_e}}}\frac{V}{V_\chi}d\chi,
\end{equation}
 is required to be larger than about $50$ to solve the flatness and horizon problems.
Here $\chi_{{}_{N_e}}$ and $\chi_f$ are the field values at the corresponding $N_e$ and
 the final point of inflation, respectively.
In all plots in this paper,
 we take $\mathcal{P}_\zeta= (2.20\pm1.10)\times10^{-9}$ and $N_e = 60$
 for normalization of those quantities.

There are three variables $p$, $\tilde{F}$, and $\tilde{V}_C$ in the potential (\ref{eq:scalar potential_3}).
The value of $p$ is quantized such as $p=1,2,3$ in superstring theory.
Here, we use $p=2,3$ because of the stationary condition (\ref{eq:Vc_stationary condition}) in this model.
The value of $\tilde{F}$ controls the depth of the potential at the stationary point,
 which corresponds to the magnitude of flatness of the potential in the region near the origin too.
The value of $\tilde{V}_C$ is adjusted uniquely to vanish the vacuum energy at the minimum.
The overall coefficient $|W_0|$ of $V=|W_0|^2\tilde{V}$ is determined by
 the amplitude of $\mathcal{P}_\zeta$.

\begin{figure}[htbp]
 \centering
 \includegraphics[width=100mm]{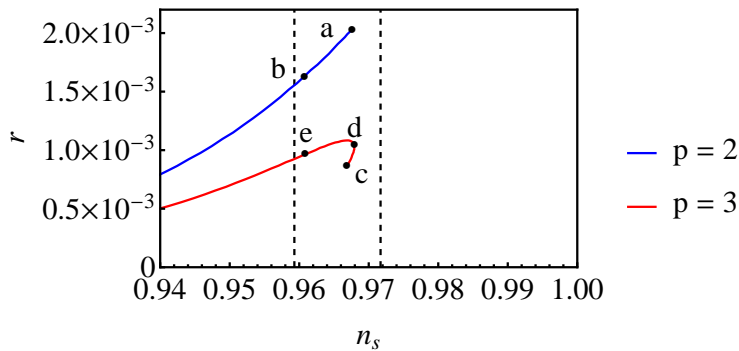}
 \caption{
$(n_s,r)$-plots for $p=2$ (blue) and $p=3$ (red). 
Two vertical dashed lines represent the lower and upper bounds for $n_s$ in the observational data (\ref{eq:Planck}).
The values of variables at sample points (a)-(e) are showed in Table \ref{tab:values_constant}.}
 \label{fig:nsr_constant}
\end{figure}

Figure~\ref{fig:nsr_constant} shows $(n_s,r)$-plots for $p=2,3$ with a parameter $\tilde{F}^2$.
The points (a) and (c) correspond to the lower limit of $\tilde{F}^2$ in inequality (\ref{eq:constraint_F}),
 and both of the curves flow to smaller $n_s$ direction as $\tilde{F}^2$ increases.
Table \ref{tab:values_constant} shows variables at each sample point (a)-(e) in Fig.~\ref{fig:nsr_constant}.
In both cases $p=2,3$, there are parameter regions of $\tilde{F}^2$ consistent
 with the observations (\ref{eq:Planck}).
At each point (a)-(e), the superpotential $|W_0|$ takes almost the same scale
 $10^{-7}M_P^3 \simeq (10^{16} \,\mathrm{GeV})^3$, and the constant potential
 $V_C\simeq(10^{15}\,\mathrm{GeV})^4$.
The mass of inflaton becomes $m^2_\chi\simeq(10^{15}\,\text{GeV})^2$. The values of $\tilde{V}_C$, $|W_0|$, and $m^2_\chi$ are analytically calculable by using approximations, and will be derived in section~\ref{sec:calculations}.

\begin{table}[htbp]
 \centering
 \begin{equation*}
  \begin{array}{|c||c|c|c|c||c|c|} \hline
\text{Point} & p & \tilde{F}^2 & \tilde{V}_C & |W_0| & n_s & r \\ \hline
\text{a} & 2 & 1.001 & 1.00\times10^3 & 2.59\times10^{-7} & 0.9676 & 2.03\times10^{-3} \\ \hline
\text{b} & 2 & 1.013 & 7.69\times10 & 8.44\times10^{-7} & 0.9607 & 1.63\times10^{-3} \\ \hline
\text{c} & 3 & 0.005 & 1.60\times10^5 & 1.33\times10^{-8} & 0.9670 & 0.87\times10^{-3} \\ \hline
\text{d} & 3 & 0.055 & 1.32\times10^4 & 1.58\times10^{-7} & 0.9679 & 1.01\times10^{-3} \\ \hline
\text{e} & 3 & 0.170 & 1.38\times10^2 & 4.82\times10^{-7} & 0.9608 & 0.96\times10^{-3} \\ \hline
  \end{array}
 \end{equation*}
 \caption{
The values of variables at each point (a)-(e) in Figure~\ref{fig:nsr_constant}.
The former four variables are input parameters in this model, and the latter two are output quantities.
}
 \label{tab:values_constant}
\end{table}

\subsection{Attractor}
\label{sec:attractor}

We showed the numerical analysis for $p=2,3$ in Figure~\ref{fig:nsr_constant}.
In this section, we treat $p$ as not a realistic integer but a continuous parameter 
in order to examine attractor properties.

\begin{figure}[htbp]
   \centering
   \includegraphics[width=100mm]{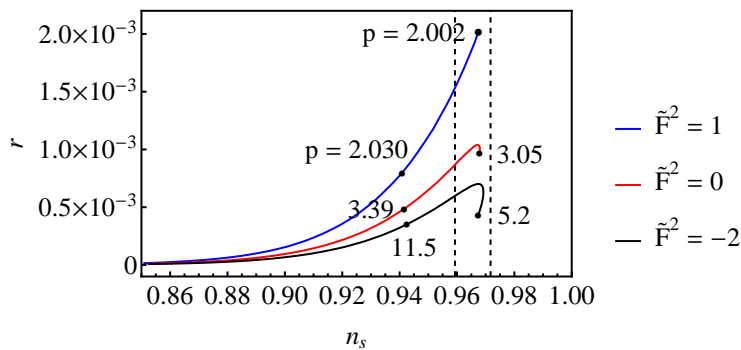}
 \caption{
$(n_s,r)$-plots for different $\tilde{F}^2$. 
The blue, red and black curves represent $\tilde{F}^2=1,0,-2$, respectively.
The number written next to each point on curves represents the value of $p$ at the point.
}
 \label{fig:nsr_attractor_1}
\end{figure}

Figure~\ref{fig:nsr_attractor_1} shows $(n_s,r)$-plots by varying $p$ continuously
 for several values of $\tilde{F}^2$.
The value of $p$ is bounded by the stationary condition (\ref{eq:constraint_F}).
Each curve flows to smaller $n_s$ direction as $p$ increases. 
It is found that the spectral index $n_s$ becomes realistic values  around $3-p \approx \tilde F^2$. 

\begin{figure}[htbp]
   \centering
   \includegraphics[width=100mm]{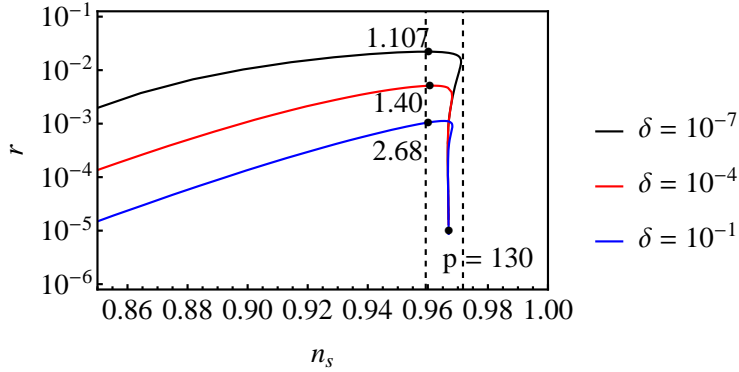}
 \caption{$(n_s,r)$-plots for different $\delta$. 
The number written next to each point on curves represents the value of $p$ at the point.
At the region near $n_s\simeq0.967$, three curves almost coincide with each other,
 and they take almost the same $r$ at $p\gtrsim100$.
}
 \label{fig:nsr_attractor_2}
\end{figure}

It is convenient to define $\delta$ by
\begin{equation}
 \delta \equiv \tilde{F}^2 + p -3. \label{eq:def_delta}
\end{equation}
The stationary point (\ref{eq:Vc_stationary point_chi}) and the stationary condition (\ref{eq:constraint_F}) can be rewritten as
\begin{equation}
 \phi^2=1-\frac{\delta}{p-1},\quad 0<\delta<p-1 \label{eq:constraint_delta},
\end{equation}
respectively.
Smaller $\delta$, which makes the stationary point closer to the pole, realizes a deeper vacuum. 
Figure~\ref{fig:nsr_attractor_2} shows $(n_s,r)$-plots by varying $p$ continuously for several $\delta$.
For a sufficiently large $p$, the spectral index $n_s$ becomes realistic.
These values do not change and approach to  $n_s\simeq0.967$ as $p$ increases.
Also, as $p$ increases, $r$ becomes smaller.
That is, we find an attractor behavior.
The point at $p=130$ is not a fixed point. For instance, $r=\mathcal{O}(10^{-6})$ can be realized by $p=1000$ on each curve.
These results show that $\tilde F^2 \approx 3-p$ is favorable.
In the limit $\tilde F^2 \rightarrow 3-p$ and $\delta \rightarrow 0$, the potential becomes sharp around the minimum.
Note that although the limit $\delta \rightarrow 0$ is favorable, point (e) with $\delta = 0.17$ in Table 1 
also leads to $n_s$ consistent with the Planck result.

\subsection{Analytical calculations}
\label{sec:calculations}

We can examine analytically the attractor behavior shown in section~\ref{sec:attractor} . 
At first, we show the attractor behavior.
Next, we derive the values of quantities mentioned in the last paragraph in section~\ref{sec:models}.

The potential (\ref{eq:scalar potential_3}) is rewritten by
\begin{equation}
 \tilde{V}=-p\cosh^{2(p-1)}\left(\frac{\chi}{\sqrt{2p}}\right)+\delta\cosh^{2p}\left(\frac{\chi}{\sqrt{2p}}\right)+\tilde{V}_C ,
\end{equation}
by using $\delta$ defined in Eq.~(\ref{eq:def_delta}). 
We can neglect the factor $\exp(-\chi/\sqrt{2p})$ in $\cosh(\chi/\sqrt{2p})$
 in the region related to inflation, and we obtain
\begin{equation}
 \begin{split}
 &\tilde{V}=-\frac{1}{2^{2p}}\left(4pe^{-\sqrt{\frac{2}{p}}\chi}-\delta\right)e^{\sqrt{2p}\chi}+\tilde{V}_C, \\
 &\tilde{V}_\chi=-\frac{\sqrt{2p}}{2^{2p}}\left[4(p-1)e^{-\sqrt{\frac{2}{p}}\chi}-\delta\right]e^{\sqrt{2p}\chi}, \\
 &\tilde{V}_{\chi\chi}=-\frac{1}{2^{2p-1}}\left[4(p-1)^2e^{-\sqrt{\frac{2}{p}}\chi}-p\delta\right]e^{\sqrt{2p}\chi}.
 \end{split}
\label{eq:potential_derivative}
\end{equation}
Being aware of $\tilde{V}\simeq\tilde{V}_C$ during inflation and with Eqs.~(\ref{eq:potential_derivative}),
 the number of e-folds is rewritten as
\begin{equation}
 N_e=\int^{\chi_{{}_{N_e}}}_{\chi_f}\frac{V}{V_\chi} d\chi
\simeq-\frac{2^{2p-2}\tilde{V}_C}{\sqrt{2p}(p-1)}\int^{\chi_{{}_{N_e}}}_{\chi_f}d\chi\,e^{-\sqrt{2p}(1-\frac{1}{p})\chi}\left[1-\frac{\delta}{4(p-1)}e^{\sqrt{\frac{2}{p}}\chi}\right]^{-1}.
\label{eq:e-foldsing_1}
\end{equation}
In the limit of $\delta\to0$ (or $p\to\infty$), 
 the terms independent of $\delta$ in Eq.~(\ref{eq:e-foldsing_1}) are dominant and we obtain
\begin{equation}
 N_e=\frac{2^{2p-3}\tilde{V}_C}{(p-1)^2}e^{-\sqrt{2p}(1-\frac{1}{p})\chi_{{}_{N_e}}}. \label{eq:e-foldsing_2}
\end{equation}
Note that the factor $e^{-\sqrt{2p}(1-\frac{1}{p})\chi_f}$ is negligible because $\chi_f$ is sufficiently large.
By using (\ref{eq:e-foldsing_2}) with $\delta =0$,
 slow-roll parameters (\ref{eq:def_slow-roll parameters}) are recast as
\begin{equation}
 \left.\epsilon\right|_{\chi=\chi_{{}_{N_e}}}=\frac{p}{4(p-1)^2N_e^2},\quad\left.\eta\right|_{\chi=\chi_{{}_{N_e}}}=-\frac{1}{N_e}.
\label{eq:epsilon_eta_approximation}
\end{equation}
Observable quantities (\ref{eq:def_observable quantities}) are expressed with $N_e$ as
\begin{align}
 n_s=1-\frac{2}{N_e}-\frac{3p}{2(p-1)^2N_e^2}, \quad r=\frac{4p}{(p-1)^2N_e^2}. 
\label{eq:nsr_approximation}
\end{align}

With the realistic value of $p$, e.g. $p=3$, and $N_e=60$, we obtain
\begin{align}
 n_s=0.9664, \quad r=0.83\times10^{-3}. \label{eq:nsr_example1}
\end{align}
These are consistent with the point (c) in Table \ref{tab:values_constant}.
For a larger $p$, instead we obtain
\begin{align}
 n_s\simeq0.9667-\frac{1}{p}\times10^{-3}, \quad r\simeq\frac{1}{p}\times10^{-3}. \label{eq:nsr_example2}
\end{align}
The second term in $n_s$ can be neglected for sufficiently large $p$, and we obtain $n_s\simeq0.9667$.
Eq.~(\ref{eq:nsr_example2}) is consistent with $(n_s,r)=(0.9669,\,1.01\times10^{-5})$ at $p=130$
 in Figure~\ref{fig:nsr_attractor_2}, and shows attractor behavior.

Next, we estimate the values of $\tilde{V}_C$, $|W_0|$, and $m^2_\chi$.
The stationary point $\chi_0$ is found from $\tilde{V}_{\chi}=0$ in Eqs.~(\ref{eq:potential_derivative}) as
\begin{equation}
 e^{-\sqrt{\frac{2}{p}}\chi_0}=\frac{\delta}{4(p-1)}.
\end{equation}
Since the constant potential $\tilde{V}_C$ is determined to realize vanishing vacuum energy $\tilde{V}|_{\chi=\chi_0}=0$, we obtain
\begin{equation}
 \tilde{V}_C=\left(\frac{p-1}{\delta}\right)^{p-1}.
\label{eq:VC_approximation}
\end{equation}
The size of superpotential $|W_0|$ is determined by the amplitude of the power spectrum $ \mathcal{P}_\zeta$.
By using Eqs.~(\ref{eq:Planck}), (\ref{eq:epsilon_eta_approximation}) and (\ref{eq:VC_approximation}),
 we obtain
\begin{equation}
 |W_0|^2=24\pi^2 \mathcal{P}_\zeta\cdot\frac{p}{4(p-1)^2N_e^2}\cdot\left(\frac{\delta}{p-1}\right)^{p-1}
\simeq\frac{3p\delta^{p-1}}{(p-1)^{p+1}}\times10^{-11}.
\label{eq:W0_approximation}
\end{equation}
The mass of inflation $m^2_\chi$ is obtained by 
\begin{align}
 m^2_\chi&=\left.V_{\chi\chi}\right|_{\chi=\chi_0}=|W_0|^2\left.\tilde{V}_{\chi\chi}\right|_{\chi=\chi_0}=\frac{6\pi^2 \mathcal{P}_\zeta p\delta^{p-1}}{N_e^2(p-1)^{p+1}}\cdot\frac{2(p-1)^p}{\delta^{p-1}} \nonumber \\
 &\simeq\frac{6p}{p-1}\times10^{-11}.
\end{align}

\section{$\phi$ dependent superpotential}
\label{sec:superpotential}

While we have considered a $\phi$-independent superpotential in previous sections for simplicity,
in general superpotential is a function of $\phi$ as well.
One expects that sufficiently small $\phi$-dependence in $W$ does not affect inflation
 in the previous sections.
In this section, we include $\phi$-dependence in $W$ and analyze inflation behavior of the scalar potential.

As an example for $\phi$-dependence, we take the following superpotential
\begin{equation}
W=W_0(1+a\phi^2+b\phi^3), \label{eq:superpotential}
\end{equation}
where $a$ and $b$ are assumed to be real constants.
Then, in the scalar potential, the CP of $\phi$ field can be violated.
Compared to a $\phi$-independent $W$ case, $V$ depends on the phase direction of $\phi$, $\theta$,
 and at the potential minimum its direction becomes massive.

We define the canonically normalized field of the imaginally direction as $\lambda$.
We have confirmed that, at the field region of very early stage of inflation
 ($\chi_\text{ini} < \chi_{N_e} , \,\lambda=0$),
 $V_\lambda=0$ and $|V_{\chi\chi}|>|V_{\lambda\lambda}|\times\mathcal{O}(10)$
 with $V_{\chi\chi},\,V_{\lambda\lambda}<0$.
Thus, the evolution of $\lambda$ direction is negligible
 and we can regard the $\chi$ direction as the inflaton.

\begin{figure}[htbp]
   \centering
   \includegraphics[width=120mm]{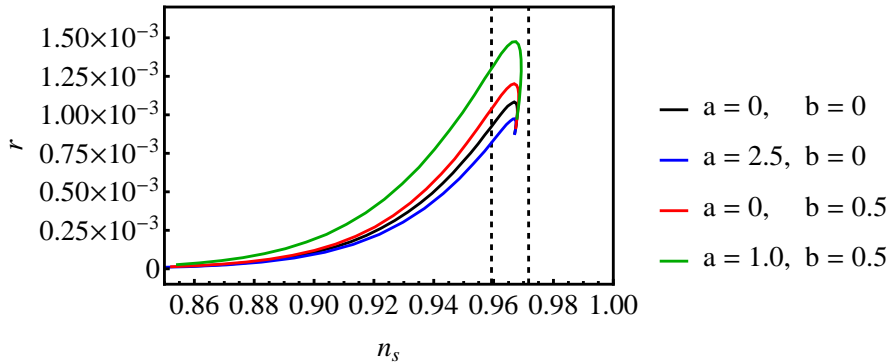}
 \caption{$(n_s,r)$-plots and different $a$ and $b$ with $p=3$.
The Black curve is the same as Figure~\ref{fig:nsr_constant}.
}
 \label{fig:nsr_superpotential}
\end{figure}

Figure~\ref{fig:nsr_superpotential} shows $(n_s,r)$-plots for different fixed $a$ and $b$ with $p=3$ and the parameter $\tilde{F}^2$.
The end point of the right hand side of each curve represents the lower bound for
 $\tilde{F}^2$ by inequality (\ref{eq:constraint_F}), i.e. $\delta \rightarrow 0$,
 and each curve flows from the point to the outside direction of the figure as $\tilde{F}^2$ increases.
We can see that $\phi$-dependence in the superpotential does not spoil inflation even if $a,b=\mathcal{O}(1)$.

\section{$D$-term potential}
\label{sec:D-term potential}

In section~\ref{sec:constant potential}, we assume the constant $V_C$ in the potential (\ref{eq:scalar potential_1}).
Since the role of $V_C$ is to uplift the potential to give vanishing vacuum energy,
 we can consider another uplifting term depending on the field $\chi$.
In this section, we show one example, that is
the D-term scalar potential written by $V_D = \frac12 g^2 D^2$ with $g$ being a gauge coupling.
For earlier works on such attempts in a differnet context, see e.g., Refs.~\cite{Burgess:2003ic,Achucarro:2006zf}
 for D-term uplifting and Ref.~\cite{deCarlos:2007sow} for inflation.
When this is independent of $\phi$, it corresponds to the constant $V_C$.
Here, we consider the case that the gauge coupling $g$ depends on $\phi$ as
\begin{equation}
g^{-2} = g^{-2}_0 + \beta \ln\left(\frac{\phi}{M_P}\right).
\end{equation}  
The second term depends on $\phi$ and such a dependence would appear
 when charged matter fields become massive by its expectation value of $\phi$.
$g_0$ and $\beta$ denote the $\phi$-independent part of the gauge coupling
 and the corresponding beta function coefficient.
Also we assume that $D$ itself is independent of $\phi$ and just a constant.
Then, the full scalar potential is written by 
\begin{equation}
 V=e^K\left[K^{\phi\bar{\phi}}\left|D_\phi W\right|^2-3\left|W\right|^2+F^2\right]+V_D.
\end{equation}
After redefinition of parameters, we simplify the D-term potential as
\begin{equation}
 V_D=\frac{D^2}{1+\alpha\ln\phi}. \label{eq:VD}
\end{equation}
Since a sufficiently small $\alpha$ makes this $D$-term sufficiently flat in order not to disturb inflation,
 $V_D$ plays a role of those of $V_C$ in the potential (\ref{eq:scalar potential_2}).

\begin{figure}[htbp]
 \centering
 \begin{tabular}{c}
  \begin{minipage}{0.5\hsize}
   \centering
   \includegraphics[width=80mm]{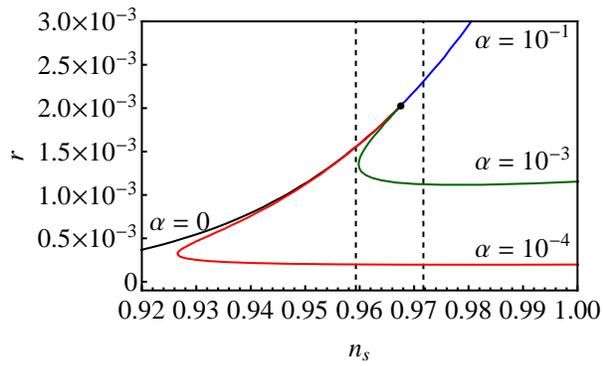}
   \subcaption{$p=2$, $\alpha>0$} \label{fig:nsr_Dterm_p2PositiveAlpha}
  \end{minipage}
  \begin{minipage}{0.5\hsize}
   \centering
   \includegraphics[width=80mm]{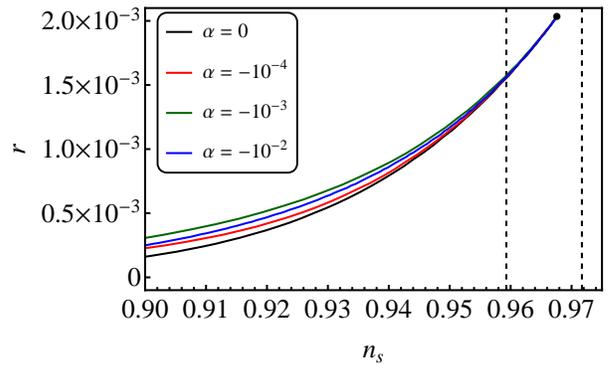}
   \subcaption{$p=2$, $\alpha<0$} \label{fig:nsr_Dterm_p2NegativeAlpha}
  \end{minipage} \\ \\
  \begin{minipage}{0.5\hsize}
   \centering
   \includegraphics[width=80mm]{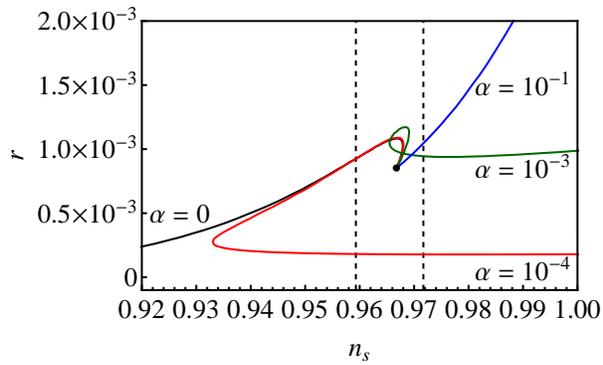}
   \subcaption{$p=3$, $\alpha>0$} \label{fig:nsr_Dterm_p3PositiveAlpha}
  \end{minipage}
  \begin{minipage}{0.5\hsize}
   \centering
   \includegraphics[width=80mm]{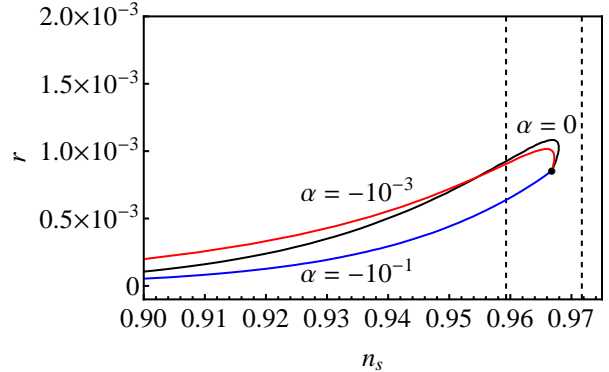}
   \subcaption{$p=3$, $\alpha<0$} \label{fig:nsr_Dterm_p3NegativeAlpha}
  \end{minipage}
 \end{tabular}
 \caption{$(n_s,r)$-plots for $p=2,3$ with positive and negative $\alpha$. 
The black curve in each figure is the same as Figure~\ref{fig:nsr_constant}.
}
 \label{fig:nsr_Dterm}
\end{figure}

Figure~\ref{fig:nsr_Dterm} shows $(n_s,r)$-plots for different fixed $\alpha$ with the parameter $\tilde{F}^2$.
The point at the center of the figure corresponds to $\delta \rightarrow 0$,
 and each curve flows from the point to the outside direction of the figure as $\tilde{F}^2$ increases.
Thus, even if we add $\phi$-dependent $V_D$, there is no change in the limit $\delta \rightarrow 0$ from the results in section~\ref{sec:constant potential},
 and the region with enough small $\delta$ is favorable for any value of $p$ and $\alpha$.

We can discuss more generic form $V(\phi)$.
Judging from the above results, we can expect that successful inflationary potential is realized with not only adding the constant term $V_C$, but also adding sufficiently flat potential around enough small $\delta$.

\section{Conclusion}

\label{sec:Conclusion}

We have studied inflationary dynamics and the resultant density perturbations
 in supergravity models, whose K\"{a}hler metric has a pole and
 scalar potential also has the pole at the same field value.
We found that successful inflation is possible by appropriate choice of parameters.
In particular, the parameter region around $\delta \approx 0$ is favorable.
For $\phi$-dependent $V_D(\phi)$ term in scalar potential instead of constant uplifting,
 this property is unaffected and the parameter region around $\delta \approx 0$ is still favorable.
 
We have also studied the attractor behavior by changing $p$ as a continuous parameter.
The parameter region around $\delta \approx 0$ is favorable, again.
For a large enough $p$, as $p$ increases,
 $n_s$ does not change approaching to $n_s \simeq 0.967$ and only $r$ decreases.

\subsection*{Acknowledgement}

The authors would like to thank Naoya Omoto for useful discussions.
This work is supported in part by the Grant-in-Aid for Scientific Research 
 No.~26247042 (T.K), No.~26400243 (O.S) from the Ministry of Education, Culture, Sports,
 Science and Technology in Japan.

\end{document}